\documentclass[aps,prl,twocolumn,superscriptaddress,amsmath,amssymb,floatfix]{revtex4-2}

\usepackage{graphicx}
\usepackage{bm}
\usepackage{xcolor}
\usepackage[colorlinks=true,linkcolor=blue,citecolor=blue,urlcolor=blue]{hyperref}
\usepackage{float}

\newcommand{\gV}{\mathrm{gV}}
\newcommand{\KL}{\mathrm{KL}}

\begin{document}

\title{Earthquake magnitude dependence: Scale invariance through local broken scaling}

\author{Didier Sornette}
\affiliation{Institute of Risk Analysis, Prediction and Management, Southern University of Science and Technology, Shenzhen, China}
\author{Giuseppe Petrillo}
\affiliation{CNRS, ENS de Lyon, LPENSL, UMR 5672, 69342 Lyon, France}

\date{28 November 2026}

\begin{abstract}
We uncover a predictive signature of genealogical memory in earthquake magnitudes hidden beneath the apparent simplicity of the Gutenberg-Richter law, opening a distinct avenue for forecasting large earthquakes. Unlike conventional triggering models based on magnitude independence, Vere-Jones's self-similar cascade reconciles scale invariance with locality through a two-branch magnitude distribution for triggered earthquakes, with a kink at the magnitude of the triggering earthquake. Conditional on earthquake times and locations in Southern California, this law outperforms a smooth magnitude-dependent law by 24.8 nats in 2019-2022 and, without refitting, by 8.8 nats in 2023-2025, corresponding to cumulative predictive likelihood ratios of \(6\times10^{10}\) and \(6.6\times10^3\), respectively. Shifting the kink by just \(\pm0.25\) magnitude units eliminates more than half the forecasting gain in 2019-2022, supporting the specific break in scaling predicted by the theory rather than merely the existence of magnitude dependence.
\end{abstract}

\maketitle

Can the history of seismicity tell us not only when and
where the next earthquake is likely to occur, but also
how large it will be? The prevailing paradigm in
statistical seismology offers a largely negative answer.
While earthquakes exhibit remarkable clustering in space
and time, their magnitudes are generally assumed to be
independent of past seismicity
\cite{Ogata88,Felzer04}. This assumption, much stronger
than the absence of magnitude correlations, imposes a
fundamental restriction on earthquake predictability.
An increase in seismic activity raises the probability
of earthquakes of all magnitudes, but carries no
additional information on the size of an individual
forthcoming earthquake. Large earthquakes are thus
regarded as rare draws from the same magnitude
distribution as small ones, and their predictability
within standard triggering models derives entirely
from fluctuations in the seismicity rate \cite{HS03}.
Whether earthquakes possess an intrinsic magnitude
memory therefore touches upon one of the most
fundamental questions in earthquake physics \cite{PZ22}.

Remarkably, the assumption of magnitude independence
also creates a fundamental difficulty in reconciling
earthquake triggering with scale invariance.
The Gutenberg--Richter (GR) law,
$P(m)\propto e^{-\beta m}$, a power law in seismic
energy spanning many decades \cite{GR44}, is a paradigm
of scale invariance. Yet the epidemic-type aftershock
sequence (ETAS) model, which underlies operational
aftershock forecasting \cite{Ogata88}, does not share
this property. An earthquake of magnitude $M$ triggers
on average $\propto e^{\alpha M}$ direct offspring,
whose magnitudes follow the GR law independently of
$M$. In a stationary GR population, mothers in
$[M,M+dM]$ therefore contribute to the triggering
of earthquakes of any given magnitude at a rate
$\propto e^{-(\beta-\alpha)M}dM$. For $\alpha<\beta$,
as usually observed \cite{Helm03}, the total triggering
rate diverges at small magnitudes, an ultraviolet (UV)
catastrophe, whereas for $\alpha>\beta$, it diverges at large
magnitudes, an infrared (IR) catastrophe. In the
scale-invariant limit $\alpha=\beta$, all magnitude
intervals contribute equally, leading to logarithmic
divergences at both ends. The UV divergence is
conventionally regularized by introducing a minimum
triggering magnitude $m_0$, which controls the
branching ratio and the partition between triggered
and background seismicity \cite{SW05a,SW05b}, yet
lies well below the completeness of the best catalogs,
likely below magnitude zero \cite{SZL25}.
Independently of these triggering divergences, the
unbounded GR law with the observed exponent implies
an infinite mean seismic moment (IR divergence) and, under the
conventional magnitude--energy scaling, an infinite
mean energy release. Physical consistency therefore
requires a modification of the upper tail
\cite{SS99}.

Generalizing L\'evy-stable
random measures, Vere-Jones \cite{VJ05}
shows that exact self-similarity requires
that the mean number of direct offspring of magnitude $u$ of a mother earthquake of magnitude $M$ per unit
magnitude depend only on the magnitude difference $u-M$ and can be written as
$\kappa e^{-\beta(u-M)}\phi(u-M)$.
In a GR population with exponent $\beta$, mothers in
$[M,M+dM]$ contribute to the triggering of earthquakes
of magnitude $m$ at a rate
$\propto e^{-\beta m}\phi(m-M)dM$.
The integrated triggering rate is finite without UV
or IR cutoffs if and only if
$\int\phi(x)\,dx<\infty$.
The fundamental requirement is thus locality of
triggering in magnitude space, which necessarily
introduces a dependence between mother and offspring
magnitudes. The scale-invariant limit of ETAS,
$\alpha=\beta$, corresponds precisely to the opposite
situation, $\phi=\mathrm{const}$, in which triggering
is completely nonlocal in magnitude.

The simplest way to reconcile locality of triggering
with GR scaling is to preserve the power law form on
either side of the natural reference magnitude provided
by the triggering earthquake itself. This leads to
two GR branches, with exponent $\beta_-<\beta$ below
$M$ and $\beta_+>\beta$ above it. Vere-Jones takes
$\phi(x)=e^{-\delta|x|}$, yielding
$\beta_\pm=\beta\pm\delta$ \cite{VJ05,SS05}.
Here lies the paradox: a local break in scaling
is the very mechanism that preserves global scale
invariance. Each earthquake introduces its own
crossover magnitude, but no absolute magnitude scale
is introduced. The mechanism is reminiscent of
the locality in scale of Kolmogorov's turbulent
cascade, which renders inertial-range statistics
asymptotically independent of the forcing and
dissipation scales \cite{K41,Eyink05}. Earthquakes
of a given magnitude receive most of their triggering
contributions from mothers within a characteristic
magnitude difference $\sim1/\delta$ \cite{NOS22}.
Moreover, with the conventional energy scaling
$E(u)\propto10^{3u/2}$, the energy released by the
direct offspring of a given mother has a finite
expectation without either magnitude cutoff whenever
$\beta_-<(3/2)\ln10<\beta_+$.
Finiteness of the total energy release of the
earthquake population remains a separate constraint
on its marginal magnitude distribution \cite{SS99}.

This construction changes the perspective on
earthquake magnitude dependence. Far from being
an empirical correction to the GR law or a
secondary feature of earthquake interactions,
magnitude memory emerges as a natural ingredient
of a nonsingular self-similar cascade. It also
makes a remarkably specific prediction: the
signature of this memory is not merely a
correlation between earthquake magnitudes, but
a change of scaling exponent that follows the
magnitude of each triggering earthquake. Such a
dependence modifies the probability that an
earthquake will trigger a larger one \cite{NOS22}, with direct
implications for the interpretation of foreshocks
and the predictability of large earthquakes.

Detecting this signature is particularly challenging.
The marginal magnitude distribution can be
blind to the two underlying exponents $\beta_\pm$
\cite{SS05}, while the genealogy of earthquake triggering is not
directly observable but can be reconstructed
probabilistically using stochastic declustering
\cite{ZOV02}. Early applications of this approach
provide evidence for magnitude dependence between
mothers and their offspring \cite{ZOV04}.
A stochastic-genealogy inversion of the California
catalog provides in-sample evidence for two branches
(central exponent $2.35$, $\delta\simeq0.74$),
although the contrast decreases to $\delta\simeq0.35$
when the weights are fixed by ETAS declustering
\cite{NOS19}. A declustering-weighted rank statistic,
not specifically designed to detect a change of
slope at the mother's magnitude, finds no magnitude
dependence in central Italy \cite{PZ23}.
More generally, correlations between nearby earthquake
magnitudes \cite{LAG24,BZG26} need not reveal a
physical memory: catalog incompleteness and persistent
variations in observation conditions can generate
apparent magnitude dependence even when the underlying
magnitudes are independent \cite{SZ26,CHM25}.
The existence, physical origin and functional form
of earthquake magnitude memory thus remain intertwined
and unresolved questions.

We approach these questions from a different
perspective, turning the theoretical prediction
into a direct forecasting experiment. We forecast
the magnitude of each Southern California earthquake
conditional on its observed time and location and on
the preceding seismic history, without assigning it
to a unique mother. All competing models share the
same catalog history and space--time triggering
weights, differing only in the magnitude law
associated with each candidate mother. Their
predictive performance is compared using the
logarithmic score \cite{Dawid84,DVJ04,GR07},
with all parameters estimated on 2016--2018 data
and subsequently held fixed. Our test thus goes
beyond asking whether earthquakes remember the
magnitudes of their mothers. It asks whether
they remember them in the specific manner predicted
by a self-similar cascade, thereby revealing a
source of predictability that is absent by
construction from standard earthquake forecasting
models.

\begin{figure}[t]
\includegraphics[width=0.9\columnwidth]{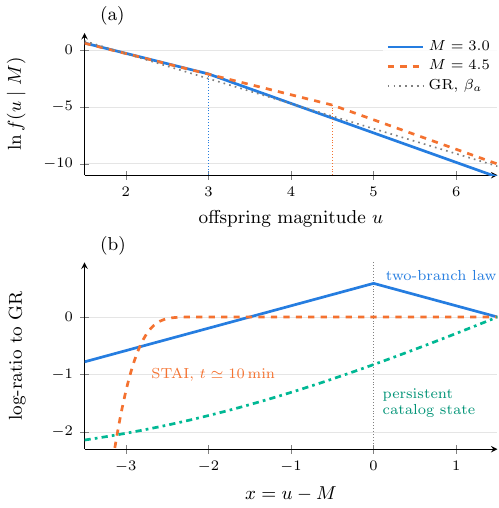}
\caption{(a) Log-density of the offspring magnitude, Eq.~\eqref{eq:gV}, for two mothers ($m^*=1.5$, $\beta_a=2.2$, $\delta=0.39$): the change of slope follows the mother's magnitude. (b) Log-ratio to the GR law for a mother of magnitude $M=5.0$. The two-branch law gives a tent with apex at $x=0$ [Eq.~\eqref{eq:tent}]. Short-term incompleteness (detection curve centered at $M-2.88$, width $0.2$) gives a monotone roll-over far below the mother. A persistent two-state catalog state rescaling magnitudes by $1$ or $1.4$ gives a smooth convex curve. Curves are shifted to vanish at $x=1.5$, and parameters of the null curves are illustrative.}
\label{fig:law}
\end{figure}

\textit{Two-branch law.} For a mother of magnitude $M\ge m^*$, the density of a direct offspring of magnitude $u\ge m^*$ is [Fig.~\ref{fig:law}(a)]
\begin{equation}
f(u\mid M)=\frac{1}{Z(a)}e^{-\beta_-(u-m^*)-2\delta\,(u-M)_+},
\label{eq:gV}
\end{equation}
with $a=M-m^*$, $\beta_\pm=\beta_a\pm\delta$ ($\beta_a=\beta$ in the self-similar cascade), $(x)_+=\max(x,0)$, and
\begin{equation}
Z(a)=\frac{1}{\beta_-} (1-e^{-\beta_-a}) + \frac{1}{\beta_+} e^{-\beta_-a}~.
\label{eq:Z}
\end{equation}
Here $m^*$ is the lower edge of the magnitude range of the catalog of mothers and offspring ($m^*=1.495$ for a cut at $1.5$ with magnitudes reported to $0.01$).
The log-density is continuous, with a slope that jumps from $-\beta_-$ to $-\beta_+$ at $u=M$. The limit $\delta=0$ recovers the GR law. Equation~\eqref{eq:gV} applies directly to the truncated catalog without requiring knowledge of the physical
minimum magnitude \cite{SM}.  Relative to a GR law with exponent $\beta_a$,
\begin{equation}
\ln f(u\mid M)-\ln\!\left[\beta_ae^{-\beta_a(u-m^*)}\right]=c(a)-\delta\,|u-M|,
\label{eq:tent}
\end{equation}
with $c(a)=\delta a-\ln[\beta_aZ(a)]$: a symmetric tent peaked at $u=M$ [Fig.~\ref{fig:law}(b)],
expressing the exponential decay of triggering locality
with magnitude separation $|u-M|$.

\textit{Conditional forecast.} The marked intensity of the catalog reads
\begin{equation}
\lambda(t,\bm x,m\mid\mathcal H_t)=\mu(\bm x)f_b(m)+\sum_{i:t_i<t}q_i(t,\bm x)\,f_a(m\mid M_i),
\label{eq:intensity}
\end{equation}
where $q_i$ is the occurrence rate contributed by the earlier event $i$, and $f_b$ and $f_a$ are normalized background and offspring magnitude densities. Conditioning on an event at $(t_j,\bm x_j)$ cancels the total rate and gives the predictive magnitude law
\begin{equation}
\begin{aligned}
p_j(m)&=w_{0j}f_b(m)+\sum_iw_{ij}f_a(m\mid M_i),\\
w_{ij}&=\frac{q_{ij}}{\mu_j+\sum_kq_{kj}},
\end{aligned}
\label{eq:mixture}
\end{equation}
with $w_{0j}=\mu_j/(\mu_j+\sum_kq_{kj})$. The weights do not use the target magnitude $m_j$. Each target is scored by its predictive bin probability $P_j(m\mid m\ge C)$, the probability that the forecast assigns to the reported bin $B(m)=[m-0.005,m+0.005)$, the whole mixture being conditioned on the reporting cut $C$:
\begin{equation}
P_j(m\mid m\ge C)=\int_{B(m)}p_j(u) du\Big/\int_{C-0.005}^{\infty}p_j(u)du.
\label{eq:binmass}
\end{equation}
For a competitor $A$, the gain of the two-branch law over the evaluation set is
\begin{equation}
G_A=\sum_jd_j,\qquad d_j=\ln P_{\gV,j}(m_j)-\ln P_{A,j}(m_j),
\label{eq:gain}
\end{equation}
in nats. If $A$, with the same weights, is the true conditional law, then
\begin{equation}
\mathbb E_A\!\left[d_j\mid\mathcal H_{t_j}\right]=-\KL\!\left(P_{A,j}\Vert P_{\gV,j}\right)\le0 .
\label{eq:KL}
\end{equation}

\textit{Observational nulls.} As a convenient approximation to short-term aftershock
incompleteness (STAI), we adopt the phenomenological
completeness threshold \cite{HKJ06,Kag04}
\begin{equation}
M_c(t)=M-4.5-0.75\log_{10}t ,
\label{eq:STAI}
\end{equation}
where $t$ is the elapsed time in days since an earthquake
of magnitude $M$.  Figure~\ref{fig:law}(b) contrasts the signature of the
two-branch law with those of two mechanisms that can
generate apparent magnitude dependence.
In the STAI model, the detection-induced roll-over
follows the mother's magnitude $M$ but lies
$\Delta(t)=4.5+0.75\log_{10}t$ magnitude units below it,
with $\Delta\ge2.1$ for $t\ge1$ min \cite{SM}.
More fundamentally, for a fixed observation state,
if detection probability is a non-decreasing function
of magnitude, so is the ratio of the observed density
to the underlying GR density \cite{SM}.
Incomplete detection thus cannot reproduce the
descending branch of the tent \eqref{eq:tent}.

A more subtle mechanism arises from persistent variations
in detection efficiency, magnitude calibration
\cite{SZ26,TWH10}, or the local $b$ value.
We represent these variations by a latent catalog state
$S$, conditional on which magnitudes are independent
draws from a state-dependent density $f_S$.
Earlier magnitudes then carry information about the
state $S$ rather than about the magnitude of the next
earthquake directly. The predictive density is
\begin{equation}
p_{\rm obs}(u\mid M_i,\mathcal H)
=\sum_S\Pr(S\mid M_i,\mathcal H)\,f_S(u),
\label{eq:PSMM}
\end{equation}
where the apparent dependence on $M_i$ arises entirely
through the state probabilities.
When the $f_S$ are exponential, the logarithm of this
mixture relative to any GR density is smooth and convex
\cite{SM}, in sharp contrast with the concave tent
and its corner at $u=M_i$ [Fig.~\ref{fig:law}(b)].
These observational mechanisms motivate the smooth
mother-dependent competitor introduced below and are
tested numerically in Sec.~S9 of the Supplemental Material \cite{SM}.

\textit{Data.} We use the relocated Southern California catalog \cite{HYS12} with $M_L\ge1.5$. All parameters are estimated before 2019 and then held fixed, and the occurrence weights come from a space--time ETAS model. The two-branch law competes with a local GR law, separate background and offspring GR laws, a two-branch law with a common fixed kink, and a smooth mother-dependent GR law [Eq.~\eqref{eq:smooth}]. Details are given in the End Matter.

\textit{Results.} Denoting by $C$ the minimum magnitude
of the events scored and by $H$ that of the candidate
mothers, we first consider the 9\,568 evaluation events
in 2019--2022 with $C=H=1.5$. The two-branch law gains
94.50 nats over local GR, 84.73 over separate background
and offspring GR laws, 97.95 over the fixed-kink law, and
24.84 over the smooth
mother-dependent law of Eq.~\eqref{eq:smooth}, which
already accounts for a dependence of the offspring exponent
on the mother's magnitude. The latter gain corresponds to $2.60\times10^{-3}$ nats per event and is positive in 12 of 16 quarters, with a sequence-clustered standard error of 10.46 nats for the total. The fitted half-difference between the branch exponents is $\hat\delta=0.389$ (95\% training-block interval, 0.166--0.578).

\begin{figure}[t]
\centering
\includegraphics[width=0.9\columnwidth]{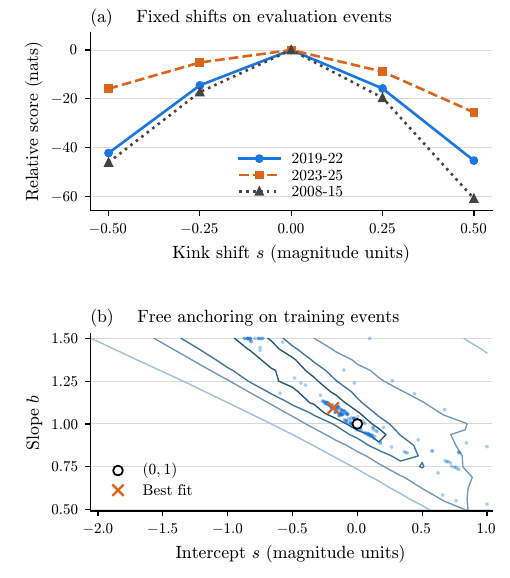}
\caption{(a) Change in evaluation log score when the kink is shifted from $M_i$ to $M_i+s$, keeping the other fitted magnitude parameters fixed. Each curve is relative to its own value at zero shift. (b) Training fit with $\kappa_i=s+bM_i$. From outside inward, contours mark losses of 8, 4, 2 and 1 nats from the best training log score, with the other magnitude parameters refitted. Dots are 200 temporal-block refits at fixed occurrence weights. The cross marks the best fit and the circle marks exact mother anchoring. The contours represent score differences, not confidence regions.}
\label{fig:anchor}
\end{figure}

With all other parameters held fixed, shifting each kink
by 0.25 magnitude units below or above its candidate
mother reduces the 2019--2022 log score by 14.5 and
15.8 nats, respectively [Fig.~\ref{fig:anchor}(a)].
To investigate whether the data favour a different
location of the kink, we generalize its position to
$\kappa_i=s+bM_i$, estimating both $s$ and $b$ from
the training data. The resulting estimates,
$\hat s=-0.185$ and $\hat b=1.094$
[Fig.~\ref{fig:anchor}(b)], are close to the theoretical
prediction $(s,b)=(0,1)$. The marginal 95\% block-bootstrap
intervals, $[-0.85,0.79]$ and $[0.71,1.50]$, respectively,
indicate substantial uncertainty, the latter reaching
the upper search bound. More importantly, allowing these
two additional parameters improves the evaluation log
score by only 2.00 nats, from 24.84 to 26.84 nats relative
to the smooth model. Thus, the theoretically predicted
location of the kink captures nearly all the predictive
gain of the freely positioned kink. Since both formulations
assume an abrupt change of slope, this comparison
supports its location but does not establish its sharpness.

The gain persists across variations in magnitude thresholds,
source weights and local $b$ values, and neither of the two
simulated observational nulls reproduces the primary gain
(see \textit{Alternative explanations} in the End Matter and
Table~\ref{tab:gains}). At $C\ge2.5$, the data are too sparse
to discriminate the laws.

\begin{figure}[t]
\centering
\includegraphics[width=0.9\columnwidth]{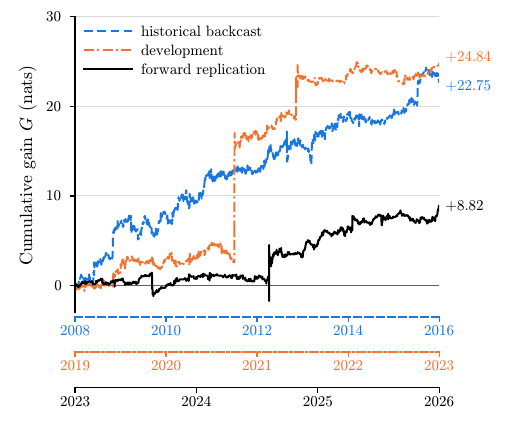}
\caption{Cumulative log-score gain of the mother-anchored law over the smooth law. Each curve starts a new sum on its own time axis. All three panels use the same pre-2019 fits. The two large steps in ``development'' (orange dotted-dashed line) coincide with the 2020 and 2021 Brawley swarms.}
\label{fig:replication}
\end{figure}

The gain is concentrated partly in two swarms, but remains
positive when their spatial cell is excluded. With the same
pre-2019 fits, the historical backcast and the previously unused
2023--2025 forward evaluation give gains of $22.75\pm7.70$
and $8.82\pm4.56$ nats, respectively (sequence standard
errors), with the unshifted kink again giving the best score
among the five positions tested [Figs.~\ref{fig:replication}
and \ref{fig:anchor}(a)]. See \textit{Swarms and replication}
in the End Matter for details.
Our test concerns the magnitude law, not exact self-similarity of the full cascade (see \textit{Open questions} in the End Matter).

\textit{Consequences.} Nandan \textit{et al.} \cite{NOS22} show that combining the two-branch law with a magnitude-dependent Omori kernel improves seismicity forecasts. Their preferred model predicts that large earthquakes, unlike small ones, are preferentially triggered by comparably large, well-recorded events. The two-branch law also increases the probability of a direct offspring exceeding a sufficiently large mother, despite its steeper upper branch, and enhances the amplification of sequences following large events. These effects can remain hidden in the marginal GR law (see \textit{Consequences of magnitude memory} and Fig.~\ref{fig:conseq} in the End Matter).

\textit{Discussion.} By holding the space--time triggering weights fixed, our forecasting experiment isolates the magnitude law. The gains over the smooth mother-dependent law and their replication without refitting provide direct out-of-sample evidence for the specific magnitude memory predicted by Vere-Jones. Beyond fluctuations in the seismicity rate, the history of seismicity thus carries information about earthquake sizes, opening a new avenue for forecasting large earthquakes. 

\textit{Acknowledgment.} G.P. acknowledges support from the CNRS at ENS de Lyon through the DyD\'eCo project. 
DS acknowledges partial support from the National Natural Science Foundation of China (Grant Nos. T2350710802 and 42604118) and the Center for Computational Science and Engineering at Southern University of Science and Technology.

\clearpage
\section*{End Matter}
\textit{Data and implementation.} We use the waveform-relocated catalog of Hauksson, Yang, and Shearer \cite{HYS12} (1981--2025 release), with 2016--2022 for development, in the rectangle $33^\circ$-$35.5^\circ$N, $115.5^\circ$-$119.5^\circ$W. We retain local magnitudes $M_L$ and quality-selected locations \cite{SM}, leaving 15\,138 development-period events with $M_L\ge1.5$. The first 300 days serve only as history. All parameters are estimated on events before 1 January 2019. The development evaluation covers 2019--2022. We also score 2008--2015 retrospectively and 2023--2025 as an untouched forward evaluation, updating the history event by event without re-estimation. 

The rate contributed by event $i$ at a target $(t_j,\bm x_j)$ is
$q_{ij}=Ke^{\alpha(M_i-1.5)}h(t_j-t_i)\,k_i(\bm x_j-\bm x_i),$
with a normalized Omori kernel $h(\tau)\propto(\tau+c)^{-p}$ on $0<\tau\le300$ d and a spatial kernel $k_i(\bm r)\propto D_i^\rho(|\bm r|^2+D_i)^{-1-\rho}$ truncated at 50 km, where $D_i=de^{\gamma(M_i-1.5)}$. With $\gamma=0.8$ and $\rho=1$ fixed, the estimates are $K=0.297$, $\alpha=0.871$, $c=2.85\times10^{-4}$ d, $p=0.936$, and $d=0.005$ km$^2$. The background has a rate in each of forty $0.5^\circ$ cells and a kernel-smoothed density within each cell, since ignoring spatial variations of the background rate biases the branching ratio upward \cite{NROS21,NOWS17}.
We fit the productivity exponent $\alpha$ freely and do not constrain the branching ratio to be less than one. Local GR exponents $\tilde\beta_c$ are estimated in each cell from the training magnitudes with a regional regularization, and the background and offspring exponents are $e^{\eta_b}\tilde\beta_c$ and $\beta_a=e^{\eta_a}\tilde\beta_c$, with a common $\delta$. 

The competitors are (i) a local GR law, (ii) separate background and offspring GR laws, (iii) a two-branch law whose kink sits at a common fitted magnitude $\kappa$, and (iv) a smooth mother-dependent GR law with offspring exponent
\begin{equation}
\beta_s(M_i)=e^{\eta_a}\tilde\beta_c\exp[g(M_i-2.5)],
\label{eq:smooth}
\end{equation}
which has three magnitude parameters, like the two-branch law.

\textit{Alternative explanations.} The gain over the smooth
mother-dependent law remains positive for all nine combinations
of three reporting cuts ($1.5$, $1.7$, $2.0$) and three
source-weight specifications (isotropic, locally elliptical,
and doubled spatial scales). Estimating local $b$ values on a
$0.1^\circ$ rather than a $0.5^\circ$ grid gives 25.45 instead
of 24.84 nats at $C=1.5$. Even when the competing GR exponent
varies independently across half-unit bins of candidate-mother
magnitude, the two-branch law gains 20.45 nats. 

Allowing
independent slope changes below and above the candidate mother
also gives a positive steepening of the upper branch,
$\hat\delta_+=0.219$ (95\% block interval, 0.025--0.371).
Raising the scoring threshold after nearby earthquakes according
to Eq.~\eqref{eq:STAI} leaves 9\,421 targets and gives
$G_{\rm smooth}=16.00$ nats and $\hat\delta=0.374$. Two
rate-based cuts adapted from Ref.~\cite{Hai16} reduce the gain to 7.55 and 8.25 nats.
Table~\ref{tab:gains} summarizes the sensitivity to magnitude
thresholds and exclusions after larger events. At $C\ge2.5$,
there are too few events to discriminate the laws. Neither
observational null reproduces the primary 24.84-nat gain in
128 simulations per null (Supplemental Material, Sec.~S9).

\begin{table}[ht]
\centering
\caption{Out-of-sample log-score gains of the mother-anchored two-branch law over the fixed-kink and smooth mother-dependent laws, with isotropic occurrence weights and evaluation in 2019--2022. The parameters are estimated before 2019. $C$ and $H$ denote the minimum magnitudes of scored events and candidate mothers, respectively. Unless otherwise specified, $H=1.5$. For $C=H=2.0$, all models are refitted on the training data. For $C=2.5$ and $3.0$, the entries give ranges over $H\in\{1.5,2.0,2.5\}$. Rows (a) and (b) exclude targets within one hour and one day, respectively, of an earlier $M_L\ge4$ event within 20 km. The final column gives $G_{\rm smooth}/N$ in units of $10^{-3}$ nats.}
\label{tab:gains}
\begin{ruledtabular}
\setlength{\tabcolsep}{2pt}
\begin{tabular}{lrrrr}
Selection & $N$ & $G_{\rm fixed}$ & $G_{\rm smooth}$ & per event\\
 & & (nats) & (nats) & ($10^{-3}$)\\
\hline
$C=1.5$ & 9568 & 97.95 & 24.84 & 2.60\\
$C=1.7$ & 6216 & 58.01 & 19.50 & 3.14\\
$C=2.0$ & 3226 & 30.44 & 11.43 & 3.54\\
$C=H=2.0$ & 3226 & 18.13 & 12.15 & 3.77\\
$C=1.5$ (a) & 9020 & 54.17 & 14.87 & 1.65\\
$C=H=2.0$ (b) & 2466 & 6.22 & 3.42 & 1.39\\
$C=2.5$ & 1110 & $-2.1$ to $3.5$ & $-1.3$ to $1.8$ & \\
$C=3.0$ & 386 & $-2.2$ to $-0.9$ & $-0.1$ to $2.1$ & \\
\end{tabular}
\end{ruledtabular}
\end{table}

\textit{Swarms and replication.} The Westmorland and Calipatria
swarms account for the two largest upward steps in the
development curve [Fig.~\ref{fig:replication}]. They contain
17.4\% of the targets but supply 59.8\% of the gain. Removing
their entire spatial cell from the history, estimation and
evaluation, and refitting the models, still gives 8.68 nats
over 6\,310 targets. 

To assess the reproducibility of the gain,
we apply the frozen pre-2019 fits to two disjoint periods.
The historical backcast [Fig.~\ref{fig:replication}(a)] gives
$22.75\pm7.70$ nats over 15\,841 events. The untouched forward
replication in 2023--2025 [Fig.~\ref{fig:replication}(c)] gives
$8.82\pm4.56$ nats over 5\,986 previously untested events
(sequence standard errors, one-sided $p=0.0265$ for the forward
period). In both periods, the theoretically predicted kink
position, $\kappa_i=M_i$, again gives the highest score among
the five shifts tested [Fig.~\ref{fig:anchor}(a)]. Further
statistical details are given in Sec.~S9 of the Supplemental
Material.

\textit{Consequences of magnitude memory.} Three implications of the two-branch law provide independent tests of magnitude memory.

\begin{figure}[ht]
\centering
\includegraphics[width=\columnwidth]{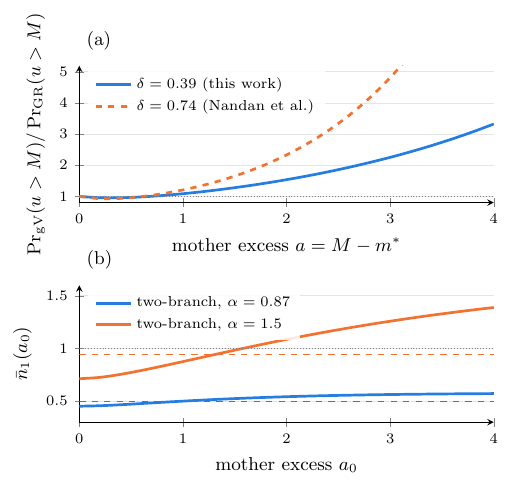}
\caption{Consequences of the two-branch law ($\beta_a=2.2$, $\delta=0.39$ unless stated). (a) Probability that a direct offspring exceeds its mother, relative to the GR law, Eq.~\eqref{eq:exceed} ($\beta_a=2.35$ for $\delta=0.74$). (b) Branching ratio of the first generation of a sequence, Eq.~\eqref{eq:n1}, with $K=0.3$. Dashed lines denote the GR values. For $\alpha=1.5$, the sequences of mothers with $a\gtrsim1.8$ are transiently supercritical while the GR value stays below one.}
\label{fig:conseq}
\end{figure}

\textit{Larger offspring.} For $a=M-m^*$, the probability that a direct offspring exceeds its mother is
\begin{equation}
\Pr(u>M\mid M)=\frac{e^{-\beta_-a}}{\beta_+Z(a)}\ \longrightarrow\ \frac{\beta_-}{\beta_+}\,e^{-\beta_-a},
\label{eq:exceed}
\end{equation}
which decays more slowly than the GR value $e^{-\beta_aa}$. For $\beta_a=2.2$ and $\delta=0.39$, the probability ratio is $1.54$ at $a=2$ and $2.25$ at $a=3$ [Fig.~\ref{fig:conseq}(a)]. Despite the steeper upper branch, the shallow lower branch shifts probability toward the mother's magnitude, increasing the chance that a large earthquake triggers a larger one \cite{NOS22}, with implications for foreshock identification.

\textit{Cascade amplification.} With productivity $\nu(M)=Ke^{\alpha(M-m^*)}$, the mean number of direct offspring per first-generation aftershock of a mother of magnitude $M$ (the first-generation branching ratio) is, for $\alpha<\beta_+$,
\begin{equation}
\bar n_1(a)=\frac{K}{Z(a)}\left[\frac{1-e^{-(\beta_--\alpha)a}}{\beta_--\alpha}+\frac{e^{(\alpha-\beta_-)a}}{\beta_+-\alpha}\right].
\label{eq:n1}
\end{equation}
For a single GR law with $\alpha<\beta_a$, $\bar n_1=K\beta_a/(\beta_a-\alpha)$ does not depend on $a$. For $\alpha>0$ and $\delta>0$, Eq.~\eqref{eq:gV} instead gives an increasing function of $a$, with the following large-$a$ limits:
\begin{equation}
\bar n_1(a)\ \sim\ \begin{cases}
K\beta_-/(\beta_--\alpha), & \alpha<\beta_-,\\[2pt]
K\beta_-a, & \alpha=\beta_-,\\[2pt]
A e^{(\alpha-\beta_-)a}, & \beta_-<\alpha<\beta_+,
\end{cases}
\label{eq:n1lim}
\end{equation}
where $A=K\beta_-[(\alpha-\beta_-)^{-1}+(\beta_+-\alpha)^{-1}]$ in the last case. Aftershocks of larger events are thus more productive, and sequences can be transiently supercritical ($\bar n_1>1$) even when the regional branching ratio is below one [Fig.~\ref{fig:conseq}(b)]. Stationarity with unbounded magnitudes requires $\alpha\le\beta_-$, a more restrictive bound than in ETAS with $\beta_a=\beta$. Local supercritical episodes remain compatible with finite realizations \cite{HS02,Sor26}, and spatially resolved ETAS inversions in California yield branching ratios up to about $1.2$ \cite{NOWS17}.

\textit{Marginal magnitude law.} In the stationary regime, the rate density of all events, $\Lambda(m)=\mu f_b(m)+\int\Lambda(M)\nu(M)f(m\mid M)\,dM$, has for $\alpha<\beta_-$ the tail $\Lambda(m)\propto e^{-\beta_{\rm eff}m}$ with
\begin{equation}
\beta_{\rm eff}=\min(\beta_b,\beta_+),
\label{eq:marginal}
\end{equation}
where $\beta_b$ is the background exponent \cite{SM}. In all 40 cells of the primary fit, $\beta_b<\beta_+$, so the asymptotic GR law conceals the offspring exponents \cite{SS05}. The regional training estimate $b=0.959$ describes the observed magnitude range, not an independently measured asymptotic tail.

\textit{Open questions.} Our test neither identifies individual mother--offspring pairs (Sec.~S4 of the Supplemental Material) nor distinguishes a sharp kink from a narrow crossover. Exact self-similarity also requires the productivity law $\nu(M)\propto Z(a)e^{\beta_-a}$ [Eq.~\eqref{eq:VJprod}], yielding the asymptotic exponent $\alpha=\beta_-=\beta-\delta$ and $\alpha/\beta_b\simeq0.75$--$0.8$ for the regional magnitude-law estimates \cite{SM}. The stationary self-similar cascade requires $\beta_-<\beta_b<\beta_+$ and a branching ratio below one \cite{SS05}. For $\alpha<\beta_-$, exponential productivity instead requires a physical lower magnitude, as in ETAS \cite{SM}. The fitted $\hat\alpha=0.87$ lies below $\hat\beta_-$, but isotropic kernels can bias $\alpha$ downward \cite{GHKK22}. Stacked aftershock counts give $\alpha\simeq0.8\,\beta$ \cite{Helm03}, whereas other analyses find or assume $\alpha\simeq\beta$ \cite{Felzer04}. An independent determination using anisotropic spatial kernels is needed.

Exact self-similarity further predicts that, for $\beta_b=\beta$ and no lower cutoff, the fraction of triggering at magnitude $m$ due to mothers below $m^*$ is $\frac12e^{-\delta(m-m^*)}$ [Eq.~\eqref{eq:share}], one half at the threshold and about one third one unit above it for $\hat\delta=0.39$, a cutoff-free counterpart of the sub-threshold productivity inferred with fertility-tapered ETAS \cite{SZL25}. Physically, the mother's rupture may delimit the stressed region, within which a smaller rupture can arrest and beyond which a larger one must propagate. This interpretation and the above predictions call for independent tests. Larger catalogs are also needed to quantify gains specifically for large earthquakes.

\clearpage
\onecolumngrid
\setcounter{equation}{0}
\setcounter{figure}{0}
\renewcommand{\theequation}{S\arabic{equation}}
\renewcommand{\thefigure}{S\arabic{figure}}
\renewcommand{\theHequation}{S\arabic{equation}}
\renewcommand{\theHfigure}{S\arabic{figure}}
\begin{center}
{\large\bfseries Supplemental Material}\\[4pt]
{Earthquake magnitude dependence: Scale invariance through broken scaling}\\[2pt]
{D. Sornette and G. Petrillo}
\end{center}

\section*{S1. Independence of the conditional magnitude
law from the physical lower cutoff}

A difficulty in testing earthquake triggering models
is that the physical minimum triggering magnitude
$M_0$ may lie several magnitude units below the
catalog completeness threshold $m^*$. An attractive
property of the two-branch law is that its conditional
functional form is preserved under lower magnitude
truncation, allowing us to test magnitude dependence
without specifying $M_0$.

Consider the physical magnitude density of a direct
offspring of a mother of magnitude $M\ge M_0$,
\begin{equation}
f_{\rm phys}(u\mid M)\propto
\exp\!\left[-\beta_-(u-M_0)
-2\delta\,(u-M)_+\right],
\qquad u\ge M_0.
\label{eq:physicaldensity}
\end{equation}
Assuming completeness above $m^*>M_0$, the conditional
density of recorded direct offspring is
\begin{equation}
f(u\mid M,u\ge m^*)=
\frac{f_{\rm phys}(u\mid M)}
{\displaystyle\int_{m^*}^{\infty}
f_{\rm phys}(v\mid M)\,dv},
\qquad u\ge m^*.
\label{eq:truncation}
\end{equation}
Writing $u-M_0=(u-m^*)+(m^*-M_0)$, the factor
$e^{-\beta_-(m^*-M_0)}$ cancels between numerator
and denominator. For $M\ge m^*$, the remaining
normalization is
\begin{align}
Z(a)&=\int_0^a e^{-\beta_-z}\,dz
\nonumber\\
&\quad+e^{-\beta_-a}
\int_0^\infty e^{-\beta_+y}\,dy
\nonumber\\
&=\frac{1-e^{-\beta_-a}}{\beta_-}
+\frac{e^{-\beta_-a}}{\beta_+},
\end{align}
where $a=M-m^*$ and $z=u-m^*$. We thus recover
exactly Eqs.~\eqref{eq:gV} and \eqref{eq:Z}
of the main text, with unchanged exponents
$\beta_\pm$ and an unchanged crossover magnitude $M$.
The catalog threshold modifies the normalization
but not the underlying signature of magnitude memory.

For a mother below the completeness threshold,
$M_0\le M<m^*$, only the upper branch is accessible
to observations. Equation~\eqref{eq:truncation}
then reduces to
\begin{equation}
f(u\mid M,u\ge m^*)=
\beta_+e^{-\beta_+(u-m^*)}.
\end{equation}
The magnitude of such a mother affects its
contribution to the occurrence rate, but not the
conditional shape of its observable offspring
magnitude distribution.

This invariance under truncation provides the
basis for testing the two-branch magnitude law
using incomplete earthquake catalogs. It does
not imply that the occurrence process is independent
of $M_0$, since unrecorded mothers can contribute
to triggering and influence the estimated occurrence
weights and productivity. These distinct issues
are discussed in Sec.~S10. Similarly, when the
scoring threshold $C$ differs from the catalog
threshold, the full predictive mixture must be
renormalized according to Eq.~\eqref{eq:binmass}.

\section*{S2. Information-theoretic interpretation
and statistical significance of predictive gains}

The forecasting experiment tests whether the history
of seismicity carries information about earthquake
magnitudes beyond that captured by competing magnitude
laws. By conditioning on the occurrence of an earthquake
at its observed time and location and using identical
occurrence weights for all models, we isolate the
predictive information contained in their respective
magnitude distributions.

\textit{Sequential evaluation and propriety of the score.}
For each evaluation earthquake $j$, the predictive
probabilities $P_{\gV,j}$ and $P_{A,j}$ are constructed
from the preceding catalog history using
Eqs.~\eqref{eq:mixture} and \eqref{eq:binmass},
before revealing its magnitude $m_j$.
The parameters estimated from the training data
remain fixed, while the seismic history is updated
after each earthquake. This sequential, or prequential,
evaluation \cite{Dawid84,DVJ04} reproduces the
information available in an operational conditional
magnitude forecasting experiment.

The logarithmic score is strictly proper \cite{GR07},
meaning that its expected value is uniquely maximized
when the forecast coincides with the true probability
distribution. For any two predictive probability distributions
$P$ and $Q$ defined on the reporting bins,
\begin{equation}
\begin{aligned}
\mathbb E_P[\ln P-\ln Q]
&=\sum_m P(m)\ln\frac{P(m)}{Q(m)}\\
&=\KL(P\Vert Q)\ge0,
\end{aligned}
\label{eq:properlogscore}
\end{equation}
with equality if and only if $P=Q$.
Thus, in expectation, the true conditional magnitude
distribution receives a higher logarithmic score
than any incorrect alternative.

\textit{Testing a conditional magnitude law.}
Let $\mathcal F_j^-$ denote the information available
before revealing the target magnitude, including
the preceding seismic history $\mathcal H_{t_j}$,
the occurrence of an earthquake at $(t_j,\bm x_j)$
and the reporting threshold.
If the competing law $A$ gives the true conditional
magnitude probabilities, the expected individual
score gain $d_j$ defined by Eq.~\eqref{eq:gain} is
\begin{equation}
\mathbb E_A[d_j\mid\mathcal F_j^-]
=-\KL(P_{A,j}\Vert P_{\gV,j})\le0,
\end{equation}
which makes explicit the conditioning in
Eq.~\eqref{eq:KL} of the main text.
Conversely, if the two-branch law gives the true
conditional probabilities,
\begin{equation}
\mathbb E_{\gV}[d_j\mid\mathcal F_j^-]
=\KL(P_{\gV,j}\Vert P_{A,j})\ge0.
\end{equation}
The expected gain then measures, in nats, the
information lost by using the competing magnitude
law instead of the two-branch law.

Under the hypothesis that $A$ is correctly specified,
the centred score increments
\begin{equation}
\xi_j=d_j+\KL(P_{A,j}\Vert P_{\gV,j})
\end{equation}
satisfy $\mathbb E_A[\xi_j\mid\mathcal F_j^-]=0$.
This martingale difference property provides a
basis for statistical tests of forecast validity
\cite{SMD93}.

\textit{Comparing predictive performance.}
Neither competing magnitude law needs to coincide
exactly with the true conditional distribution.
Our empirical question is therefore more general:
does the two-branch law provide systematically
better magnitude forecasts than law $A$?
For $N$ evaluation targets, the relevant one-sided
null hypothesis of predictive performance
\cite{GW06} is
\begin{equation}
\frac{1}{N}\sum_{j=1}^N\mathbb E[d_j]\le0.
\label{eq:predictivenull}
\end{equation}
The cumulative gain $G_A=\sum_jd_j$ in
Eq.~\eqref{eq:gain} measures the realized predictive
advantage of the two-branch law over the evaluation
period. Positive values favour the two-branch law,
while $G_A/N$ measures its average gain per earthquake.
Unlike the conditional expectations above, the
realized gain is not itself a KL divergence.

\textit{Uncertainty accounting for earthquake clustering.}
Although the forecasts are constructed sequentially,
the score increments $d_j$ need not be independent.
In particular, earthquakes belonging to the same
aftershock sequence or swarm can contribute correlated
score gains. Treating individual earthquakes as
independent observations could therefore underestimate
the uncertainty of $G_A$.

We account approximately for this dependence by
constructing connected components of earthquakes,
linking pairs separated by at most 1 km and one day.
Denoting the resulting components by $\mathcal C_k$,
$k=1,\ldots,L$, and the evaluation mean by
$\bar d=G_A/N$, the sequence-clustered variance
estimator of the cumulative gain is
\begin{equation}
\widehat{\mathrm{Var}}(G_A)=
\frac{L}{L-1}
\sum_{k=1}^{L}
\left[
\sum_{j\in\mathcal C_k}(d_j-\bar d)
\right]^2.
\label{eq:scorevariance}
\end{equation}
There are $L=5\,309$ components among the 9\,568
evaluation targets in 2019--2022.
For the comparison with the smooth mother-dependent
law, this gives
$G_{\rm smooth}=24.84\pm10.46$ nats, where the
uncertainty denotes one sequence-clustered standard
error. Further comparisons and robustness tests are
reported in Sec.~S9 and Table~\ref{tab:seqchecks}.

This estimator allows dependence within components
but assumes that distinct components are approximately
independent. The components are statistical groupings,
not reconstructed earthquake genealogies.
Moreover, the uncertainties are conditional on the
parameters estimated from the training data and do
not include their estimation uncertainty.

The forecasting framework therefore provides a direct
test of the predictive information associated with
magnitude memory, without requiring the identification
of individual mother--offspring pairs. The comparison
with the smooth mother-dependent law is particularly
discriminating, since it tests the specific magnitude
dependence predicted by Vere-Jones rather than
magnitude dependence alone.

\section*{S3. Observational artifacts as alternative explanations of magnitude dependence}
\textit{Short-term incompleteness.} With Eq.~\eqref{eq:STAI} of the main text, the offset between the mother magnitude and the completeness level is $\Delta(t)=4.5+0.75\log_{10}t$ ($t$ in days):
\begin{equation}
\Delta(10\,{\rm s})=1.55,\quad \Delta(1\,{\rm min})=2.13,\quad \Delta(10\,{\rm min})=2.88,\quad \Delta(1\,{\rm h})=3.46,\quad \Delta(1\,{\rm d})=4.5 .
\end{equation}
Dynamic truncation excludes a target only if its magnitude falls below its time- and location-specific threshold, so a sufficiently large early aftershock remains eligible.

\textit{Monotone detection within a state.} Let the latent law of an offspring be $f(u)$ and let the probability of detection $q(u)$ be non-decreasing in $u$ (for any observation state). The observed law is $q(u)f(u)/\int qf$, so that
\begin{equation}
\ln p_{\rm obs}(u)-\ln f(u)=\ln q(u)+{\rm const}
\end{equation}
is non-decreasing in $u$. If $f$ is a GR law, the log-ratio of the observed law to the GR law can therefore increase or stay constant, but it cannot decrease for $u>M$. This statement compares the observed law with the same latent GR law at fixed state.

\textit{Mixtures of exponential laws are log-convex.} If, conditionally on a state $S$, the reported magnitudes are exponential with exponent $\beta_S$, as for a state-dependent rescaling of the magnitude scale \cite{SZ26} or for unresolved heterogeneity of the $b$ value, then
\begin{equation}
\ln p(u)=\ln\sum_S\pi_S\beta_Se^{-\beta_S(u-m^*)}
\end{equation}
is a log-sum-exp of affine functions of $u$, hence convex and infinitely differentiable. Its log-ratio to any GR law is convex as well. The tent is concave with a corner, so it cannot arise from such mixtures. This calculation does not cover arbitrary combinations of detection, state changes and magnitude errors. We therefore calibrate the fitted comparison under two explicit observational nulls (Sec.~S9).

\section*{S4. Empirical signature of locality in magnitude space}

The predictive gains reported in the main text provide
evidence for the two-branch magnitude law. Here, we
investigate whether its underlying physical mechanism,
an enhanced triggering contribution from earthquakes
of comparable magnitudes, leaves a more direct signature
in the statistics of magnitude differences.

For each target earthquake $j$ and each candidate
mother $i$, we consider the magnitude difference
$x_{ij}=m_j-M_i$, weighted by the corresponding
space--time triggering weight $w_{ij}$ defined in
Eq.~\eqref{eq:mixture}.
To isolate the contribution of magnitude locality
beyond a smooth dependence on the mother's magnitude,
we compare the observed weighted counts in each bin
$B$ of magnitude differences,
\begin{equation}
N_{\rm obs}(B)=
\sum_j\sum_{i:M_i\ge2.5}
w_{ij}\,\mathbf{1}_{\{m_j-M_i\in B\}},
\end{equation}
with their expectations under the full smooth
mother-dependent predictive distribution,
\begin{equation}
N_{\rm smooth}(B)=
\sum_j\sum_{i:M_i\ge2.5}
w_{ij}\,
\Pr_{\rm smooth}(u-M_i\in B\mid\mathcal H_j).
\end{equation}
The probabilities are evaluated conditional on the
observed time and location of each target and on the
catalog reporting threshold.

\begin{figure}[H]
\centering
\includegraphics[width=.60\textwidth]{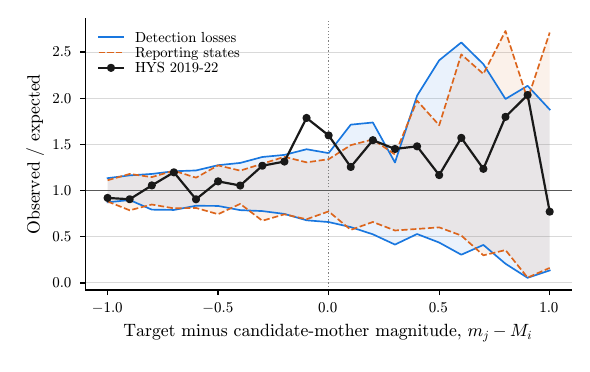}
\caption{Empirical signature of magnitude locality beyond
smooth magnitude dependence. Ratio of observed to expected
weighted counts of magnitude differences $x_{ij}=m_j-M_i$,
in bins of width 0.1, for candidate mothers with $M_i\ge2.5$.
The black curve represents the Southern California catalog
for 2019--2022. The blue solid and orange dashed curves
delimit the pointwise central 95\% ranges obtained from
32 simulated catalogs under the short-term detection-loss
and persistent reporting-state null models, respectively.
These ranges are not simultaneous confidence bands.
The horizontal reference at unity corresponds to agreement
with the smooth mother-dependent forecast. The local excess
of empirical counts near $x_{ij}=0$ supports enhanced
triggering locality in magnitude space. Statistics become
sparse at positive magnitude differences, with only
2.24 expected weighted counts at $x_{ij}=1$.}
\label{fig:empiricaltent}
\end{figure}

Figure~\ref{fig:empiricaltent} presents the ratio
$N_{\rm obs}(B)/N_{\rm smooth}(B)$ for the 2019--2022
evaluation events, together with the corresponding
ranges obtained from the two observational null models.
The bin centred at zero magnitude difference contains
26.88 weighted counts against an expectation of 16.82,
an excess of approximately 60\%. The adjacent bin on
the negative side exhibits an even larger ratio.
This local enhancement near magnitude equality is
consistent with the locality mechanism underlying
the Vere-Jones cascade.

The empirical ratio is not expected to reproduce
exactly the theoretical tent of Eq.~\eqref{eq:tent},
which refers to a single known mother and uses the
GR law as its reference distribution. The uncertainty
in earthquake genealogies, the superposition of
contributions from different candidate mothers and
the use of a smooth mother-dependent reference law
all modify the expected shape. Moreover, the small
number of large positive magnitude differences limits
the resolution of the upper branch.
The histogram therefore provides complementary
evidence for magnitude locality, while the
out-of-sample predictive comparisons and the
displacement tests of Fig.~\ref{fig:anchor}
provide a more discriminating test of the specific
Vere-Jones magnitude law.

\section*{S5. Expected predictive information from magnitude memory}
\begin{figure}[h]
\centering
\includegraphics[width=0.6\textwidth]{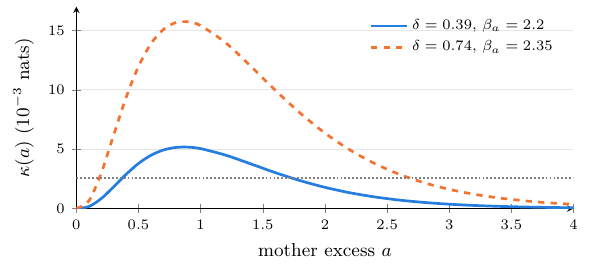}
\caption{Expected gain per direct offspring with known mother over the best single exponential law for that mother. The dotted line is the realized gain per target against the smooth competitor at $C=1.5$.}
\label{fig:budget}
\end{figure}

For a direct offspring whose mother is known, with excess $a$, the expected gain of the two-branch law over the best-fitting single exponential law for that mother is
\begin{equation}
\kappa(a)=\KL\!\left(f(\cdot\mid a)\,\Vert\,\beta^*e^{-\beta^*z}\right),\qquad \beta^*(a)=1/\mathbb E[z\mid a].
\end{equation}
$\kappa$ vanishes at $a=0$ (pure upper branch) and for $a\to\infty$ (lower branch dominant), and peaks near $a\simeq0.9$ (Fig.~\ref{fig:budget}). Averaged over the productivity-weighted distribution of mother excess, $a\sim{\rm Exp}(\beta_a-\alpha)$, it is $2.6\times10^{-3}$ nats for $(\beta_a,\delta)=(2.2,0.39)$ and $8.3\times10^{-3}$ nats for $(2.35,0.74)$. The observed gain against the smooth law, $2.6\times10^{-3}$ nats per target, is averaged over all targets, including background events and targets with diffuse weights, and might be expected to lie below this value. Part of the difference is explained by the parametric form of the smooth competitor: $\beta^*(a)$ saturates from $\beta_+$ at $a=0$ to $\beta_-$ at large $a$, and approximating it by the log-linear form of Eq.~\eqref{eq:smooth} costs about $1.0\times10^{-3}$ nats per direct offspring in the same calculation. 

Using the fitted event-specific mixtures on the reported magnitude bins gives $\sum_j\KL(P_{\gV,j}\Vert P_{{\rm smooth},j})=15.84$ nats for 2019--2022, compared with a realized gain of 24.84 nats. The former is an expectation under the fitted two-branch law, not a bound on the realized score. The concentration of $\kappa(a)$ around $a\simeq1$ may help explain why swarms, rich in mothers well above the cut, carry much of the gain.

\section*{S6. Magnitude-dependent cascade amplification}

In standard ETAS, large earthquakes trigger more direct
offspring than small ones, but the expected productivity
of these offspring is independent of the magnitude of
their mother. The two-branch magnitude law introduces
a qualitatively different mechanism: large earthquakes
tend to generate more productive offspring, enhancing
the subsequent amplification of their sequences.

Consider an initiating earthquake of magnitude
$M_0=m^*+a_0$. We define the first-generation branching
ratio $\bar n_1(a_0)$ as the expected number of
second-generation offspring per first-generation
offspring, or equivalently the ratio of the expected
sizes of these two generations. With the productivity
law $\nu(z)=Ke^{\alpha z}$, where $z=u-m^*$ is the
offspring magnitude excess, we obtain
\begin{equation}
\bar n_1(a_0)=
\int_0^\infty f(z\mid a_0)\,Ke^{\alpha z}\,dz.
\end{equation}
Using the two-branch law \eqref{eq:gV} and splitting
the integral at $z=a_0$ gives
\begin{equation}
\bar n_1(a_0)=\frac{K}{Z(a_0)}
\left[
\int_0^{a_0}e^{-(\beta_--\alpha)z}dz
+e^{-\beta_-a_0}
\int_{a_0}^\infty
e^{-\beta_+(z-a_0)+\alpha z}dz
\right],
\end{equation}
which yields Eq.~\eqref{eq:n1} of the End Matter
for $\alpha<\beta_+$. For $\alpha\ge\beta_+$,
the second integral diverges, even for a finite
initiating magnitude.

For $\alpha>0$ and $\delta>0$, $\bar n_1(a_0)$
increases monotonically with $a_0$, as the offspring
distribution shifts probability towards larger
magnitudes. At $a_0=0$, it takes its minimum value,
$K\beta_+/(\beta_+-\alpha)$.
For large $a_0$, three regimes emerge,
\begin{equation}
\bar n_1(a_0)\sim
\begin{cases}
\displaystyle\frac{K\beta_-}{\beta_--\alpha},
& \alpha<\beta_-,\\[6pt]
K\beta_-a_0,
& \alpha=\beta_-,\\[4pt]
\displaystyle
K\beta_-\left(
\frac{1}{\alpha-\beta_-}
+\frac{1}{\beta_+-\alpha}
\right)e^{(\alpha-\beta_-)a_0},
& \beta_-<\alpha<\beta_+.
\end{cases}
\end{equation}
In contrast, the magnitude-independent GR law gives
the constant branching ratio
$n_{\rm GR}=K\beta_a/(\beta_a-\alpha)$,
provided $\alpha<\beta_a$.

For $K=0.3$, $\beta_a=2.2$, and $\delta=0.39$,
the GR branching ratio is 0.50 for $\alpha=0.87$
and 0.94 for $\alpha=1.5$. With the two-branch law,
$\bar n_1(a_0)$ increases from 0.45 to 0.58 and
from 0.71 to 1.75, respectively, as $a_0$ increases
from zero to infinity.

The second example illustrates a remarkable consequence
of magnitude memory: the expected population of the
second generation can exceed that of the first for
sufficiently large initiating earthquakes, even when
the corresponding magnitude-independent cascade is
subcritical. This generational amplification does not
necessarily imply sustained supercriticality of the
entire cascade, but provides a mechanism for enhanced
transient seismic activity following large earthquakes.

Magnitude dependence thus affects not only the sizes
of individual earthquakes but also the collective
dynamics of their triggering sequences. An independent test of this prediction would estimate
$\bar n_1(a_0)$ as a function of the initiating
earthquake's magnitude using stochastic reconstruction
of earthquake genealogies \cite{ZOV02,ZOV04,Sor26}.

\section*{S7. Why the marginal Gutenberg--Richter law
conceals magnitude dependence}

Can a cascade with a strong dependence between mother
and offspring magnitudes reproduce the same marginal
GR law as a cascade with independent magnitudes?
This question is central to understanding why magnitude
memory is so difficult to detect from conventional
earthquake statistics.

In a stationary regime, the marginal rate density
satisfies the self-consistency equation
\begin{equation}
\Lambda(m)=\mu f_b(m)+
\int_{m^*}^{\infty}\Lambda(M)\nu(M)f(m\mid M)\,dM,
\end{equation}
where the first term represents background seismicity
and the second accounts for triggering by earthquakes
of all magnitudes.

Writing $x=m-m^*$ and $y=M-m^*$, and using the
two-branch law \eqref{eq:gV} with productivity
$\nu(M)=Ke^{\alpha y}$, we decompose the triggered
contribution according to whether the mother is smaller
or larger than the offspring,
\begin{align}
T_<(x)&=Ke^{-\beta_+x}
\int_0^x
\frac{\Lambda(m^*+y)}{Z(y)}
e^{(\alpha+2\delta)y}\,dy,\\
T_>(x)&=Ke^{-\beta_-x}
\int_x^\infty
\frac{\Lambda(m^*+y)}{Z(y)}
e^{\alpha y}\,dy.
\end{align}
Assuming an asymptotic marginal GR law,
$\Lambda(m)\propto e^{-\beta m}$, and using
$Z(y)\to1/\beta_-$ for large $y$, we obtain, for
$\beta>\alpha$,
\begin{equation}
T_>(x)\propto e^{-(\beta_-+\beta-\alpha)x},
\end{equation}
while the contribution from smaller mothers behaves as
\begin{equation}
T_<(x)\propto
\begin{cases}
e^{-\beta_+x}, & \beta>\alpha+2\delta,\\
xe^{-\beta_+x}, & \beta=\alpha+2\delta,\\
e^{-(\beta_-+\beta-\alpha)x},
& \beta<\alpha+2\delta.
\end{cases}
\end{equation}
Since the slowest exponential decay controls the
asymptotic magnitude distribution, self-consistency
requires
\begin{equation}
\beta=\min\!\left(
\beta_b,\ \beta_+,\ \beta_-+\beta-\alpha
\right).
\end{equation}

Three regimes emerge. For $\alpha<\beta_-$, the triggering
exponent $\beta_-+\beta-\alpha$ is strictly larger than
$\beta$ and therefore cannot determine the asymptotic
decay of the marginal magnitude distribution. Assuming
that a stationary solution exists, self-consistency thus
requires $\beta_{\rm eff}=\min(\beta_b,\beta_+)$,
as given by Eq.~\eqref{eq:marginal}.
Remarkably, when $\beta_b<\beta_+$, the background
GR exponent controls the marginal distribution,
irrespective of the contrast between the two offspring
branches \cite{SS05}. For $\alpha>\beta_-$, triggering generates
a heavier tail than the assumed marginal distribution,
precluding a stationary GR solution with unbounded
magnitudes. At the boundary $\alpha=\beta_-$,
both triggering contributions scale as $e^{-\beta m}$
for $\beta_-<\beta<\beta_+$. The balance of exponents
alone no longer determines the stationary distribution,
and one must also match amplitudes. For the exactly self-similar cascade, matching the
amplitudes yields the stationarity condition
$n(\beta_b)<1$, where $n(\beta_b)$ is the eigenvalue
of the triggering operator given by
Eq.~\eqref{eq:VJeigenvalue} in Sec.~S10.

We illustrate these results by solving the stationary
equation numerically on a magnitude grid
($m^*=0$, step $0.005$, upper limit $14$,
$K=0.3$, $\alpha=0.87$, $\beta_a=2.2$ and
$\delta=0.39$). For $\beta_b=2.2$ and $\beta_b=3.0$,
the predicted asymptotic exponents are $2.20$ and
$2.59$, respectively, while the fitted exponents over
the magnitude interval $[6,10]$ are $2.23$ and $2.61$.
For $\alpha=1.9>\beta_-=1.81$, the numerical scheme
yields no positive stationary solution. The finite-range
estimates also illustrate the slow convergence towards
the asymptotic GR exponent.

The essential result is that the marginal GR law
can conceal the magnitude dependence governing the
cascade. Two earthquake populations obeying the same
GR law may have very different conditional probabilities
of producing large earthquakes. Testing magnitude
independence therefore requires probing the conditional
rather than the marginal magnitude distribution,
as in the forecasting experiment of the Letter.

\section*{S8. Enhanced probability of triggering a larger earthquake}

One of the most consequential implications of magnitude
memory concerns the probability that an earthquake turns
out to be a foreshock of a larger event. At first sight,
the steeper upper branch of the two-branch law suggests
that earthquakes larger than their mothers should be
less probable than under the standard GR law.
Remarkably, the opposite holds for sufficiently large
mothers, because the shallower lower branch redistributes
probability towards the mother's magnitude.

From Eq.~\eqref{eq:gV}, the probability that a direct
offspring exceeds its mother's magnitude $M$ is
\begin{equation}
\Pr(u>M\mid M)
=\frac{e^{-\beta_-a}}{Z(a)}
\int_0^\infty e^{-\beta_+y}dy
=\frac{e^{-\beta_-a}}{\beta_+Z(a)},
\end{equation}
where $a=M-m^*$. For large $a$, $Z(a)\to1/\beta_-$,
yielding Eq.~\eqref{eq:exceed}.
Relative to the magnitude-independent GR prediction
$e^{-\beta_a a}$, the enhancement factor is
\begin{equation}
R(a)=\frac{e^{\delta a}}{\beta_+Z(a)}
\simeq\frac{\beta_-}{\beta_+}e^{\delta a},
\qquad a\gg1.
\end{equation}
The relative enhancement thus increases exponentially
with the mother's magnitude. For
$(\beta_a,\delta)=(2.2,0.39)$, the enhancement factors
are $1.09$, $1.54$ and $2.25$ at $a=1$, 2 and 3,
respectively. For $(2.35,0.74)$, corresponding to the
estimates of Nandan \textit{et al.} \cite{NOS19},
they reach $1.21$, $2.33$ and $4.82$.

Accounting for the productivity of the mother,
the expected number of direct offspring exceeding
its magnitude scales asymptotically as
$e^{(\alpha-\beta_-)a}$, rather than
$e^{(\alpha-\beta_a)a}$ under the standard GR law.
Magnitude memory therefore modifies not only the
conditional distribution of earthquake sizes but also
the expected frequency of triggering sequences in
which a large earthquake produces a still larger one.
This identifies a source of information relevant
to forecasting large earthquakes that is absent
from magnitude-independent triggering models.

\section*{S9. Statistical details and robustness checks}
\textit{Data selection.} \textit{Data selection.} We retain local magnitudes $M_L$
and earthquakes relocated using GrowClust, a hierarchical
waveform-based relocation algorithm \cite{TS17}, with
estimated horizontal location errors below 1 km.
For earthquakes located by other methods, we require
horizontal errors below 2 km and travel-time residuals
below 0.4 s.

\textit{Evaluation.} The quarter-resampling ranges use 2\,000 resamples of the 16 evaluation quarters, with the fitted models held fixed. The one-sided sign test for 12 positive quarters out of 16 gives $p=2517/65536=0.038$, whereas against the fixed kink, 13 of 16 gives $p=0.011$. At $C=H=2.0$, the mean gain per event against the smooth law is $3.77\times10^{-3}$ (range $[0.26,6.29]\times10^{-3}$) for all targets, $2.04\times10^{-3}$ ($[-0.05,4.09]\times10^{-3}$) outside the first hour after $M_L\ge4$ events, and $1.39\times10^{-3}$ ($[-0.58,3.32]\times10^{-3}$) outside the first day. At $C=1.5$, the 548 targets of the first hour after $M_L\ge4$ events (5.7\%) supply 40\% of the gain against the smooth law.

\textit{Pair statistics.} A two-branch fit to all forward pairs within 5 km and 7 days (two-sided magnitude differences, 25\,000 of 154\,413 eligible pairs) gives effective exponents $1.90$ and $2.19$, i.e., a half-contrast of $0.15$. These pairs are dominated by multiplicity: 80.4\% of them have their first event in the Brawley cell, and their number exceeds the median of matched ETAS simulations by a factor $6.45$. A near-far contrast of first larger partners, which uses at most one partner per event, gives a contrast of the sign opposite to that expected from the two-branch law. These pair statistics are not used for inference in the main text, as the branch decomposition below addresses the upper branch directly.

\textit{Location and uncertainty of the kink.} We profile the training log score over $-4\le s\le1$ and $0\le b\le1.5$ for the sharp hinge $\kappa_i=s+bM_i$, refitting the magnitude parameters at each position. We resample the nine three-month training blocks 200 times, recalculating the local GR slopes and refitting the magnitude laws at fixed occurrence weights. The reported intervals are marginal percentile intervals, not a joint confidence region for the anchor. At $b=0$ the intercept range places the kink below our reporting cut, so the freely fitted common-kink model in Table~\ref{tab:gains}, not the edge of the affine search, is our observable fixed-kink control.

\textit{Additional magnitude controls.}
In the two-branch model, the central offspring exponent
$\beta_a=e^{\eta_a}\tilde{\beta}_c$ and the background
exponent $\beta_b=e^{\eta_b}\tilde{\beta}_c$ differ from
the locally estimated GR exponent $\tilde{\beta}_c$
by multiplicative factors, with fitted log multipliers
$\eta_a=-0.0560$ and $\eta_b=0.0194$.
In the smooth mother-dependent model, the coefficient
$g=-0.1268$ in Eq.~\eqref{eq:smooth} describes the
decrease of the offspring GR exponent with increasing
mother magnitude.

The background probability $w_{0j}$ has a median of
0.0031, with lower and upper quartiles of $1.30\times10^{-5}$
and 0.994, reflecting the coexistence of predominantly
triggered and predominantly background events.

To test whether the predictive advantage of the
two-branch law results from an overly restrictive
parametrization of the smooth competitor, we consider
two more flexible alternatives. Allowing the offspring
GR exponent to vary independently in each 0.5-unit
bin of candidate-mother magnitude still yields a gain
of 20.45 nats in favour of the two-branch law.
An adaptive alternative updates a local Gamma
distribution of the GR exponent using the magnitude
excesses of earlier neighbouring earthquakes.
Integrating the GR law over this distribution of
exponents yields a power law predictive density
(the Lomax distribution), accounting for local
fluctuations and uncertainty in the GR exponent
without introducing explicit genealogical memory.
The two-branch law
outperforms this adaptive model by 52.23 nats when
the weights are scaled to the effective number of
neighbours and by 65.57 nats with unscaled weights.
These comparisons indicate that the predictive gain
of the two-branch law is not simply a consequence
of insufficient flexibility in the competing
magnitude distributions.

For the asymmetric fit, we tie the central offspring slope to the background slope and fit separate changes below and above $M_i$. We obtain $\hat\delta_-=0.555$ (block interval 0.275--0.811) and $\hat\delta_+=0.219$ (0.025--0.371). For the primary symmetric-model gain against the smooth competitor, classifying targets by the single highest-weight candidate mother gives $+30.87$ nats above its magnitude, $-10.34$ below, and $+4.31$ at equality at the catalogue precision. Four targets have no candidate mother and contribute less than 0.01 nat in absolute value. These are score contributions, not assignments of true parents.

\textit{Observation and source weights.}
To assess the sensitivity of our results to short-term
aftershock incompleteness, we maximize Eq.~\eqref{eq:STAI}
over strictly earlier events within 20 km and add
0.2 magnitude units as a safety margin. For each target,
we adopt the greater of this threshold and $C$, rounded
up to a reported magnitude bin. We refit the magnitude
laws on eligible training targets and renormalize each
predictive mixture over its target-specific magnitude
range. Of the 9\,568 evaluation targets, 9\,421 remain,
with a log-score gain over the smooth mother-dependent
law of $G_{\rm smooth}=16.00$ nats and
$\hat\delta=0.374$.

Two alternative completeness criteria based on local
seismicity rates, adapted from Ref.~\cite{Hai16},
use the ten preceding earthquakes and rate limits
of 100 and 200 events per day. They yield
$G_{\rm smooth}=7.55$ and $8.25$ nats, respectively,
with sequence-clustered standard errors of 4.48
and 4.82 nats. Although these more restrictive
completeness criteria reduce the predictive gains,
the estimated branch contrast remains remarkably
stable, with $\hat\delta=0.380$ and $0.384$,
respectively, compared with the primary estimate
of $0.389$.

\textit{Sensitivity to spatial resolution and seismic swarms.}
The parameter $d$ controls the characteristic spatial
scale $\sqrt{D_i}$ of the triggering kernel, with
$D_i=d\,e^{\gamma(M_i-1.5)}$.
To account more conservatively for location uncertainties,
we impose a lower bound on $d$ equal to twice the mean
squared horizontal location error in the training catalog.
This increases the minimum characteristic triggering length
$\sqrt d$ from 0.071 to 0.278 km, while the log-score gain
over the smooth mother-dependent law remains substantial,
$G_{\rm smooth}=20.99$ nats.

To assess the influence of the Brawley swarms, we remove
the entire spatial cell containing them (cell 7) from
the catalog history, training and evaluation sets.
Refitting the models yields $G_{\rm smooth}=8.68$ nats
over 6\,310 evaluation events, demonstrating that the
predictive gain is not confined to these episodes.

Finally, adapting the four-model comparison of
Ref.~\cite{Sor26}, we allow transient changes in
earthquake productivity, background seismicity rate,
both, or neither during the Westmorland and Calipatria
swarms. This distinction between endogenous amplification
and external forcing is particularly relevant to the
interpretation of seismic swarms \cite{LMO09}.
Our objective is to test whether the predictive
advantage of the two-branch magnitude law survives
these alternative descriptions of the occurrence process.
With the original spatial weights, the preferred models
involve a change in productivity for Westmorland and
changes in both productivity and background rate for
Calipatria, with corresponding gains in favour of the
two-branch magnitude law of 11.47 and 3.48 nats.
With the larger minimum spatial scale, the model with
unchanged productivity and background rate is preferred
for Calipatria. Thus, although the inferred triggering
mechanisms of individual swarms are sensitive to spatial
resolution, the predictive advantage of the two-branch
magnitude law persists under these alternative
specifications of the occurrence weights.

\begin{figure}[ht]
\centering
\includegraphics[width=.52\textwidth]{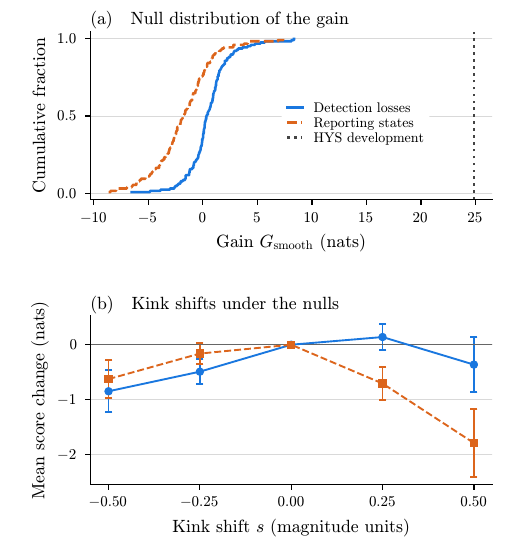}
\caption{(a) Empirical cumulative distribution of the
log-score gain $G_{\rm smooth}$, defined by
Eq.~\eqref{eq:gain}, of the two-branch law over the smooth
mother-dependent law under each observational null model.
The vertical dotted line indicates the observed 2019--2022 gain
of 24.84 nats. (b) Mean change in log score as a function
of the displacement $s$ of the kink from the candidate
mother's magnitude, relative to $s=0$, under the same
two null models. Error bars represent 1.96 Monte Carlo
standard errors of the mean across simulated catalogues.}
\label{fig:nullcal}
\end{figure}

\textit{Observational calibration.} Each of 384 seven-year synthetic catalogues is fitted separately following the primary HYS chronology \cite{HYS12}. We simulate the physical process before observation losses and estimate candidate-mother weights without using the simulated genealogy. The first no-gV generator combines local detection, short-term magnitude-dependent losses, and a 10-s blind time after equal or larger nearby events. In the second, each spatial cell switches between two reporting states (mean residence seven days), scaling the magnitude excess above 1.5 by 0.8 or 1.2. The 99th percentiles of $G$ are 8.12 and 6.89 nats, respectively (128 catalogues each), and neither set reaches the HYS development value of 24.84. The fitted median contrasts under the nulls are small but nonzero (0.046 and 0.045), and their mean shift profiles are not exactly flat [Fig.~\ref{fig:nullcal}]. Under gV with the short-term-loss detector, all 64 catalogues at each of $\delta=0.37$ and $0.74$ exceed the larger of the two simulated 99th percentiles. The median estimates are 0.391 and 0.764. These detection rates apply to these generators and sample sizes.

\textit{Frozen replication.} We use 300 days of prior history in each period, with unchanged quality filters, magnitude harmonization and fitted parameters. The historical evaluation is a backcast: the development analysis uses part of the 2008--2015 interval to screen for completeness and pair counts. The 2023--2025 magnitudes provide a previously unused forward test of gV. The gain over the smooth model is $22.75\pm7.70$ nats in the historical period and $8.82\pm4.56$ in the forward period (sequence standard errors). Adding the gains and the two period-specific variance estimates gives $31.57\pm8.94$ nats and a one-sided normal-approximation $p=0.0002$. The forward period alone gives $p=0.027$. Of its 8.82 nats, 8.03 come from outside cell 7. With all other parameters fixed, shifting the kink by $-0.25$ or $+0.25$ costs 5.14 or 8.89 nats in this period.

In 2008--2015, the gain is 4.99 nats within cell 7 and 17.76 outside it. Within cell 7 in calendar year 2012, the gain is 0.61 nats over 855 targets, but this annual subset does not isolate the Brawley swarm.

Table~\ref{tab:seqchecks} reports the uncertainties on the principal comparisons. Against the stepwise and the two adaptive slope models, the sequence standard errors are 7.17, 13.42 and 23.07 nats, respectively.
\begin{table}[ht]
\centering
\caption{Out-of-sample predictive performance of the
two-branch magnitude law relative to four competing
magnitude laws for the 2019--2022 evaluation period.
Each entry gives the cumulative log-score gain
$G_A$ defined by Eq.~\eqref{eq:gain}, in nats, together
with one sequence-clustered standard error accounting
for the dependence between earthquakes within sequences.
Positive values favour the two-branch law.
The first row corresponds to the primary analysis,
while the subsequent rows assess robustness to the
location of the kink, catalog incompleteness, spatial
location uncertainties and the exclusion of the Brawley
region. Within each row, all competing forecasts use
identical evaluation events and space--time occurrence
weights, ensuring that the gains reflect differences
in the magnitude laws rather than in the occurrence
process.}
\label{tab:seqchecks}
\begin{ruledtabular}
\begin{tabular}{lrrrr}
Analysis & Local GR & Separate GR & Fixed kink & Smooth\\
\hline
Primary & $94.50\pm31.73$ & $84.73\pm27.38$ & $97.95\pm25.23$ & $24.84\pm10.46$ \\
Free anchor & $96.50\pm32.22$ & $86.73\pm27.90$ & $99.94\pm25.76$ & $26.84\pm10.83$ \\
Dynamic threshold & $48.50\pm13.30$ & $44.70\pm11.89$ & $63.85\pm12.97$ & $16.00\pm5.77$ \\
Rate cap 100/day & $32.25\pm7.76$ & $29.55\pm7.59$ & $47.06\pm10.51$ & $7.55\pm4.48$ \\
Rate cap 200/day & $32.47\pm8.11$ & $29.90\pm8.18$ & $47.38\pm11.07$ & $8.25\pm4.82$ \\
Location-error floor & $101.42\pm36.72$ & $97.11\pm34.29$ & $118.40\pm31.62$ & $20.99\pm11.35$ \\
Without cell 7 & $19.45\pm4.61$ & $17.18\pm4.36$ & $14.26\pm4.20$ & $8.68\pm2.88$ \\
\end{tabular}
\end{ruledtabular}
\end{table}

\section*{S10. The exactly self-similar limit}
\textit{Kernel.} In Vere-Jones's cascade the mean number of direct offspring per unit magnitude of a mother of magnitude $M$ is \cite{SS05}
\begin{equation}
G(u\mid M)=\kappa\,e^{-\beta(u-M)-\delta|u-M|},
\label{eq:VJkernel}
\end{equation}
which depends on $u-M$ only: the cascade of a mother of magnitude $M+c$ is that of a mother of magnitude $M$ shifted by $c$. With $\phi(x)=e^{-\delta|x|}$, this is the form $\kappa e^{-\beta x}\phi(x)$ of the main text. Equation~\eqref{eq:VJkernel} has two GR branches, with exponent $\beta_-=\beta-\delta$ below $M$ and $\beta_+=\beta+\delta$ above it, and is the unique continuous kernel of the form $\kappa e^{-\beta x}\phi(x)$ that is a pure exponential on each side of $x=0$ with symmetric contrasts. Unequal contrasts, $\delta_-$ below and $\delta_+$ above the mother, give the asymmetric law fitted in Sec.~S9.

\textit{Productivity.} The number of direct offspring above a threshold $m^*\le M$ is
\begin{equation}
\nu(M)=\int_{m^*}^{\infty}G(u\mid M)\,du=\kappa\,Z(a)\,e^{\beta_-a},\qquad a=M-m^*,
\label{eq:VJprod}
\end{equation}
with $Z(a)$ given by Eq.~\eqref{eq:Z}, and $G/\nu$ is exactly Eq.~\eqref{eq:gV} with $\beta_a=\beta$. The occurrence model of the main text, with productivity $Ke^{\alpha a}$, reduces to this cascade for $\alpha=\beta_-$, up to the prefactor $Z(a)$, which varies between $1/\beta_+$ at $a=0$ and $1/\beta_-$ at large $a$. The fitted occurrence model ($\hat\alpha=0.87$, against $\hat\beta_-\simeq1.70$ for the regional estimates) is therefore not exactly self-similar, and the magnitude-score comparison tests the shape of Eq.~\eqref{eq:gV}, not Eq.~\eqref{eq:VJprod}. Since the marginal exponent is $\beta_b$ (below), the prediction reads $\alpha/\beta_b=\beta_-/\beta_b$, which is $0.75$ for the regional training estimates and $1-\delta/\beta=0.81$ when $\beta_b=\beta_a$.

\textit{Why $\alpha<\beta_-$ is not self-similar.} For any shift-invariant kernel $G(u\mid M)=g(u-M)$ whose lower tail is $g(x)\simeq Ae^{-\beta_-x}$ for $x\to-\infty$, the number of offspring above a fixed threshold grows as $(A/\beta_-)e^{\beta_-(M-m^*)}$. The productivity exponent and the lower-branch exponent are thus one and the same parameter, whatever the shape of $g$. For ETAS, whose offspring law has the single exponent $\beta$, the same identity gives $\alpha=\beta$, the scale-invariant case of the main text. 

With $G=Ke^{\alpha a}f(u\mid M)$ and $\alpha<\beta_-$, shifting all magnitudes by $c$ multiplies the offspring rate of every large mother by $e^{(\alpha-\beta_-)c}$: the effective coupling grows toward small magnitudes. In the parametrization $G=\kappa e^{aM-\beta u-\delta|u-M|}$ of Ref.~\cite{SS05}, $a=\alpha+\delta$, and Eq.~\eqref{eq:VJkernel} is the case $a=\beta$. Two consequences follow. (i) A GR population with exponent $\gamma$ produces first-generation offspring with exponent $\gamma+\beta_--\alpha$, so successive generations steepen, and the integral over mothers diverges at small magnitudes as soon as an exponent exceeds $\alpha+2\delta$. For the fitted values ($\hat\alpha+2\hat\delta=1.65<\hat\beta_b\simeq2.25$) this already happens in the first generation. A lower magnitude $M_0$ is therefore required, as $m_0$ is in ETAS. (ii) Observable parameters depend on $M_0$. For a mother well above $m^*$, the productivity observed above $m^*$ is $K=K_0e^{-(\beta_--\alpha)(m^*-M_0)}$, where $K_0$ is the productivity above $M_0$. Triggering by events between $M_0$ and $m^*$ depends on $M_0$ as well. At $\alpha=\beta_-$, with the productivity of Eq.~\eqref{eq:VJprod}, the dependence on $M_0$ cancels exactly, and $M_0$ can be sent to $-\infty$. We verify (i) and (ii) by numerical integration.

\textit{Stationarity.} With background magnitudes distributed as $e^{-\beta_bm}$ without lower bound, mothers in $[M,M+dM]$ contribute to the rate at magnitude $m$ a term $\propto e^{-\beta_bm}\,e^{(\beta_b-\beta)(m-M)-\delta|m-M|}\,dM$, which is integrable over $M$ if and only if $\beta_-<\beta_b<\beta_+$. The total rate is then $\Lambda=\mu/[1-n(\beta_b)]$ with \cite{SS05}
\begin{equation}
n(\beta_b)=\kappa\left[\frac{1}{\beta_b-\beta_-}+\frac{1}{\beta_+-\beta_b}\right],\qquad n(\beta)=\frac{2\kappa}{\delta},
\label{eq:VJeigenvalue}
\end{equation}
and a stationary GR rate with exponent $\beta_b$ requires $n(\beta_b)<1$. The marginal law reproduces the background exponent, in agreement with Eq.~\eqref{eq:marginal} and with Ref.~\cite{SS05}. This completes Sec.~S7 at $\alpha=\beta_-$.

\textit{Sub-threshold share of triggering.} For $\beta_b=\beta$, mothers at $M$ contribute to the triggering of events at $m$ with weight $\phi(m-M)=e^{-\delta|m-M|}$, the tent of Eq.~\eqref{eq:tent}. The share contributed by mothers below a catalog threshold $m^*\le m$ is
\begin{equation}
S(m)=\frac{\int_{-\infty}^{m^*}e^{-\delta(m-M)}dM}{\int_{-\infty}^{\infty}e^{-\delta|m-M|}dM}=\frac12\,e^{-\delta(m-m^*)},
\label{eq:share}
\end{equation}
i.e., $0.50$, $0.34$, and $0.23$ at $m-m^*=0$, $1$, and $2$ for $\delta=0.39$. For unequal contrasts, $S(m)=[\delta_-/(\delta_-+\delta_+)]\,e^{-\delta_+(m-m^*)}$. If the GR population stops at a physical lower magnitude $M_0<m^*$, the share becomes $[e^{-\delta(m-m^*)}-e^{-\delta(m-M_0)}]/[2-e^{-\delta(m-M_0)}]$. For comparison, in ETAS with $\alpha<\beta$ and minimum triggering magnitude $m_0$, the same share is $1-e^{-(\beta-\alpha)(m^*-m_0)}$: it does not decay with $m$, and it is fixed by the cutoff, tending to one as $m_0\to-\infty$.

\end{document}